\documentclass[useAMS,usenatbib]{mn2e}
\usepackage{color}
\usepackage[utf8]{inputenc}
\usepackage{amsmath}
\usepackage{wasysym}
\usepackage{amsfonts}
\usepackage{amssymb}
\usepackage{graphicx}
\title{Galactic evolution of rapid neutron capture process abundances: the inhomogeneous approach}
\author[Wehmeyer, Pignatari \& Thielemann]{B. Wehmeyer,$^1$ \thanks{benjamin.wehmeyer@unibas.ch} M. Pignatari$^{1,2}$ and F.-K. Thielemann$^1$ \\
$^1$ Univ Basel, Dept. Phys., Klingelbergstr. 82, CH-4056 Basel, Switzerland\\
$^2$ Konkoly Observatory, Research Centre for Astronomy and Earth Sciences, Hungarian Academy of Sciences,\\ Konkoly Thege Mikl\'{o}s \'{u}t 15-17, H-1121 Budapest, Hungary}
\date{Unreferreed and unedited version as per June 16, 2015 (1); In original form January 30, 2015}
\begin{document}
\maketitle
\begin{abstract}
For the origin of heavy r-process elements, different sources have been proposed, e.g., core-collapse supernovae or neutron star mergers. Old metal-poor stars carry the signature of the astrophysical source(s). Among the elements dominantly made by the r-process, europium (Eu) is relatively easy to observe. In this work we simulate the evolution of europium in our galaxy with the inhomogeneous chemical evolution model 'ICE', and compare our results with spectroscopic observations.
We test the most important parameters affecting the chemical evolution of Eu: (a) for neutron star mergers the coalescence time scale of the merger ($t_{\mathrm{coal}}$) and the probability to experience a neutron star merger event after two supernova explosions occurred and formed a double neutron star system ($P_{\mathrm{NSM}}$) and (b) for the sub-class of magneto-rotationally driven supernovae (''Jet-SNe''), their occurrence rate compared to standard supernovae ($P_{\mathrm{Jet-SN}}$).
We find that the observed [Eu/Fe] pattern in the galaxy can be reproduced by a combination of neutron star mergers and magneto-rotationally driven supernovae as r-process sources. While neutron star mergers alone seem to set in at too high metallicities, Jet-SNe provide a cure for this deficiency at low metallicities. 
Furthermore, we confirm that local inhomogeneities can explain the observed large spread in the europium abundances at low metallicities. We also predict the evolution of [O/Fe] to test whether the spread in $\alpha$-elements for inhomogeneous models agrees with observations and whether this provides constraints on supernova explosion models and their nucleosynthesis.
\end{abstract}
\section {Introduction}\label{introduction}
The rapid neutron capture process (r-process, e.g., \citealt{Thielemann11}, and references therein) is responsible for the production of about half of the heavy element abundances beyond Fe in the solar system, and of the heaviest elements like Th and U. The remaining heavy element abundances are mostly made by the slow neutron capture process (s-process, e.g., \citealt{Kaeppeler11}). Despite its relevance, the true astrophysical origin of the r-process is still under debate. Because of the larger uncertainties affecting the r-process nucleosynthesis predictions compared to the s-process in stars, the r-process isotopic contribution in the solar system has been originally identified by using the residual method, i.e., by subtracting the s-process component from the solar isotopic distribution (e.g., \citealt{Arlandini99}, \citealt{Bisterzo14}). The r-process residual abundances have been shown to be consistent in first approximation with the abundance signatures in old r-process rich metal-poor stars (at least for elements heavier than Ba, see \citealt{Travaglio04} for details), carrying the signature of the r-process nucleosynthesis in the early galaxy (e.g., \citealt{Sneden08}). For instance, Eu receives only a marginal contribution from the s-process (the s-process explains only 6 per cent of the solar Eu, while the remaining amount has an r-process origin, \citealt{Bisterzo14}), and therefore it is often used as a tracer of the r-process nucleosynthesis in stellar spectroscopic observations. One possibility to test predictions from r-process nucleosynthesis is to include the r-process stellar yields in galactic chemical evolution (GCE) simulations, and to compare the theoretical results with spectroscopic observations at different metallicities. Eu is an ideal diagnostic for these studies.
The purpose of this work is to illustrate the europium evolution throughout the evolution of our galaxy. We consider here the contribution from two sites (and their frequency) to the production of heavy r-process elements: Neutron Star Merger (NSM) and ''magneto-rotationally driven supernovae'' (hereafter referred to as ''Jet-SNe''). We will show that the combination of both sites is able to reproduce the observed europium abundance distribution of the stars of our galaxy.

Neutron star - black hole (NS-BH) mergers might have a non-negligible contribution to the r-process inventory in the galaxy. However, their relevance as astrophysical source for the r-process is controversial, since this event has not yet been observed (e.g., \citealt{Bauswein14}). These difficulties also result in an extreme divergence of the predicted galactic rate of such an event (e.g., \citealt{Postnov14}). However, it should be noticed that a contribution from NS-BH mergers has been predicted as well (e.g., \citealt{Korobkin12}, \citealt{Mennekens14}). We give an estimate of the possible effects caused by these events in Section~\ref{Conlusion and Discussion}.

Chemical evolution of galaxies has made strong advances since its early days. Initially all approaches made use of the instantaneous recycling approximation in the sense, that the ejecta of stellar end stages were immediately utilized without delay after the initial star formation, assuming that the stellar evolution time scale is short in comparison to galactic evolution. If, in addition, the instantaneous mixing approximation was applied, i.e., assuming that the ejecta were instantaneously mixed throughout the galaxy, the whole galaxy acts as a homogeneous box. Neglecting this can explain radial gradients. Further developments included infall of primordial matter into and outflow  of enriched material out of the galaxy (for a review of these early investigations see e.g., \citealt{Audouze76}). When relaxing the instantaneous recycling approximation, i.e., taking into account that (explosive) stellar ejecta enter the interstellar Medium (ISM) delayed with respect to the birth of a star by the duration of its stellar evolution, detailed predictions for the evolution of element abundances can be made. Based on nucleosynthesis predictions for stellar deaths, a number of detailed analyses have been performed, from light elements up to the Fe-group (e.g., \citealt{Timmes95}, \citealt{Goswami00}, \citealt{Matteucci01}, \citealt{Gibson03}, \citealt{Kobayashi06}, \citealt{Pagel09}, \citealt{Kobayashi12}, \citealt{Matteucci12}). Such approaches have recently also been applied to understand the enrichment of heavy elements in the galaxy (including r-process contributions) as a function of time or metallicity [Fe/H] (see e.g., \citealt{Ishimaru99}, \citealt{Travaglio99}, \citealt{DeDonder03}, \citealt{Matteucci12}, \citealt{Vangioni15}).

However, if still the instantaneous mixing approximation is applied, i.e., such ejecta are instantaneously mixed with the global ISM, no local inhomogeneities can be produced. The latter would relate to the fact that only limited amounts of the ISM are polluted by / mixed with the ejecta of each event. This effect is of essential importance, especially at low metallicities, where portions of the ISM are already polluted by stellar winds and supernovae, and others are not. In addition, different portions of the ISM are polluted by different types of events, leading to a scatter at the same metallicity, which can in fact be utilized as a constraint for these different stellar ejecta. When, however, utilizing the instantaneous mixing approximation, this leads to a unique relation between galactic evolution time and metallicity [Fe/H], i.e., any [Fe/H] can be related to a specific time in the evolution of a galaxy (while inhomogeneous mixing could experience similar [Fe/H] values in different locations of the galaxy at different times). This is especially the case in the very 
early galactic evolution ([Fe/H]$<-2.5$), when locally only a few stars (out of a whole initial mass function IMF) might have exploded and imprinted their stellar neighbourhood with their ejecta. Thus, the application of chemical evolution models which utilize the instantaneous mixing approximation is questionable for the early evolution of galaxies. 

In addition, for each [Fe/H], due to the instantaneous mixing, only a mean value of [X/Fe] (X being the element of interest to follow in chemical evolution) is obtained. Inhomogeneous mixing, however, could produce larger ratios in strongly polluted areas and smaller values in still less polluted ones. This means that the scatter in [X/Fe] at low metallicities, which might also be a helpful asset in pointing to the origin of element X, cannot be reproduced or utilized with an homogeneous treatment. In case of rare events, which – on the other hand – produce large amounts of element X in each event, this would produce a large scatter, and – if observed – could be used as a very helpful constraint to identify the production site. For these reasons, specially for the origin of r-process elements like Eu, we think that only inhomogeneous chemical evolution models should be utilized at low metallicities. The two type of rare events (i) Jet-SNe (maybe up to 1\% of all core collapse supernovae (CCSNe)) and (ii) neutron star mergers, with a similar occurrence frequency of about 1\% of all CCSNe are considered here, while regular and more frequent core collapse supernovae might at most contribute to the lighter r-process elements. The binary merger rates are estimated by \cite{vandenHeuvel96} as well as \cite{Kalogera04}. The rate of Jet-SNe is related to the fact that about 1\% of neutron stars are found with magnetic fields of the order $10^{15}$ Gauss (magnetars, see, e.g., \citealt{Kouveliotou98}, \citealt{Kramer09}).

Earlier inhomogeneous chem(odynam)ical evolution models for r-process elements like Eu have been provided by Travaglio et al. (2001, where the r-process yields were assumed to come from CCSN), \cite{Argast04} and \cite{Matteucci14} comparing neutron star mergers and core collapse supernovae, Cescutti \& Chiappini (2014, comparing NSM and Jet-SNe) Mennekens \& Vanbeveren (with NSM and NS-BH mergers), and Shen et al. and van de Voort et al. (2015, only utilizing neutron star mergers). One of the main questions here is related to the problem of reproducing [Eu/Fe] at low(est) metallicities. Cescutti \& Chiappini (2014) have shown that this is possible with Jet-SNe. \cite{Argast04} concluded that neutron star mergers cannot reproduce observations at [Fe/H]$<-2.5$, while \cite{vandeVoort15} and \citealt{Shen15} came to the opposite conclusion.

The main difference between Jet-SNe and neutron star mergers is that in one case the immediate progenitors are massive stars and the first occurrence in chemical evolution is due to the death of massive stars. In the other case the progenitors are also massive stars, leading to two supernova explosions in a binary system, which – if not disrupted – causes a binary neutron star system and a merger with a given delay time due to gravitational radiation losses. Thus, one needs to consider two aspects: (i) the two supernova explosions and the pollution of the ISM with their ejecta (for the case of NS-BH mergers see the discussion in section~\ref{Conlusion and Discussion}), and (ii) the delay time of the merger event after the formation of the binary neutron star system. Especially aspect (i) can only be treated adequately with inhomogeneous evolution models, and there an additional factor is of major importance: with how much matter the supernova ejecta mix before the neutron star mergers eject their products into the same environment.

This paper is organized as follows. In section~\ref{The model}, we introduce the model used to compute the evolution of abundances. In section~\ref{Results}, we present the influence of the different r- and non-r-process sites on the evolution. Additionally, we provide an overview why an inhomogeneous treatment of the evolution is important. In section~\ref{inhom} we discuss the impact of inhomogeneities, causing and permitting a scatter of [X/Fe] ratios at low metallicities. As a further test of the model, we discuss the fact why the large scatter of [r/Fe] observed at low metallicities is strongly reduced for $\alpha$-elements, and show how this constraints core collapse supernovae nucleosynthesis predictions, which are still not available in a self-consistent way. In section~\ref{Conlusion and Discussion}, our results are summarized and discussed.

\section{The model}\label{The model}
Recent chemodynamical galactic evolution models, like e.g., \cite{Minchev14}, \cite{vandeVoort15}, and \cite{Shen15}, can model in a self-consistent way massive mergers of galactic subsystems (causing effects like infall in simpler models), energy feedback from stellar explosions (causing effects like outflows), radial migrations in disk galaxies, mixing and diffusion of matter/ISM, and the initiation of star formation dependent on local conditions, resulting from the effects discussed above. In our present investigation we still utilize a more classical approach with a parametrized infall of primordial matter, and a Schmidt law (\citealt{Schmidt59}) for star formation. Therefore, we neglect large scale mixing effects, while we include the feedback from stellar explosions and the resulting mixing with the surrounding ISM, according to a Sedov-Taylor blast wave. In this way, the model permits to keep track of the local inhomogeneities due to different CCSN ejecta. This approach allows to grasp the main features of the impact of the first stars / stellar deaths on the evolution of the heavy element enrichment. This approximation omits other mixing effects, e.g., spiral arm mixing (on time scales of the order of $2\cdot 10^8$ years). The main focus of this work is the investigation of the chemical evolution behaviour at low metallicities, where these effects should not have occured, yet, and are therefore left out in this first order approximation.

We treat the galactic chemical evolution of europium (Eu), iron (Fe) and $\alpha$-elements (e.g., oxygen O), utilizing the established GCE code ''Inhomogeneous Chemical Evolution'' (ICE), created by \cite{Argast04}. A detailed description of the model can be found therein.

For the simulation, we set up a cube of $(2\textit{kpc})^3$ within the galaxy which is cut in $40^3$ smaller cubes representing a $(50\textit{pc})^3$ sub cube each. The evolution is followed with time-steps of 1My.
Primordial matter is assumed to fall into the simulation volume, obeying the form
\begin{equation}
\dot{ M}(t)= a \cdot t^b \cdot e^ {-t/\tau} \textit{,}
\end{equation}
which permits an initially rising and eventually exponentially declining infall rate. While $\tau$ and the total galaxy evolution time $t_{final}$ are fixed initially, the parameters $a$ and $b$ can be determined alternatively from $M_{tot}$ (the total infall mass integrated over time), defined by
\begin{equation} 
M_{tot} := \int_0^{t_{final}} a \cdot t^b \cdot e^ {-t/ \tau} \textit{,}
\end{equation}
 and the time of maximal infall $t_{max}$, given by
\begin{equation}
t_{max}:=b\cdot \tau \textit{.}
\end{equation}

See \cite{Argast04} for an extended discussion of the infall model and table~\ref{infall parameters} for the applied parameters.
\begin{table}
\begin{tabular}{|llr|}
\hline
\hline
$M_{tot}$ & Total infall mass & $10^8 \text{M}_{\odot}$ \\
$\tau$ & time scale of infall decline & $5\cdot 10^9$yrs \\
$t_{max}$ & time of the highest infall rate & $2\cdot 10^9$yrs \\
$t_{final}$ & duration of the simulation & $13.6\cdot 10^9$yrs \\
\hline
\end{tabular}
\caption{Main infall parameters. See Argast et al. (2004) for details on the parameters.}
\label{infall parameters}
\end{table}

\subsection{Treating stellar births and deaths} \label{Iteration Procedure}
The main calculation loop at each time step (1My) can be described in the following way.

\begin{enumerate} 

\item We scan all mass cells of the total volume and calculate the star formation rate per volume and time step ($10^6$ yrs) according to a Schmidt law with a density power $\alpha=1.5$ (see \citealt{Schmidt59}, \citealt{Kennicutt98}, \citealt{Larson91}). Dividing by the average stellar mass of a Salpeter IMF (power $-2.35$) provides the total number of stars per time step $n(t)$ created in the overall volume of our simulation.

\item Individual cells in which stars are formed are selected randomly until $n(t)$ is attained, but the probability is scaled with the density, which leads to the fact that patches of higher density, predominently close to supernova remnants, are chosen.

\item The mass of a newly created star is chosen randomly in the range $0.1$ to $50 M_ {\odot}$, subject to the condition that the mass distribution of all stars follows a Salpeter IMF. Consequently only cells which contain more than $50 M_{\odot}$ are selected in order to prevent a bias.

\item The newly born star inherits the composition of the ISM out of which it is formed.

\item The age of each star is monitored, in order to determine the end of its lifetime, either to form a white dwarf or experience a supernova explosion (see \ref{LMS} and \ref{HMS}). A fraction of all high mass stars ($M>8M_{\odot}$), according to the probability ($P_{Jet-SN}$), is chosen to undergo a magneto-rotationally driven supernova event (see section~\ref{Jet-SN}). Type Ia supernova events are chosen from white dwarfs according to the discussion in \ref{SNIa}. The treatment of neutron star mergers follows the description in \ref{NSM_model}.
\item The composition for the ejecta of all these events is chosen according to the discussion in \ref{Nucleosynthesis sites}. They will pollute the neighbouring ISM with their nucleosynthesis products and sweep up the material in a chemically well mixed shell. We assume that an event pollutes typically $ 5\cdot 10^4 \text{M}_{\odot} $ of surrounding ISM due to a Sedov-Taylor blast-wave of $10^{51}$erg (\citealt{Ryan96}, \citealt{Shigeyama98}). This implies that the radius of a remnant depends strongly on the local density and the density of the surrounding cells.
\item In the affected surrounding cells, stars are polluted by the matter of the previously exploded star and the event specific element yields.
\end{enumerate}

The details on the above procedure will be explained in the following.
\subsection{Nucleosynthesis sites} \label{Nucleosynthesis sites}
\subsubsection{Low (LMS) and intermediate mass stars (IMS)} \label{LMS}
Low and intermediate mass stars provide a fundamental contribution to the GCE of e.g., He, C, N, F, Na and heavy s-process elements during the asymptotic giant branch (AGB) phase. For instance, most of the C and N in the solar system were made by AGB stars (e.g., \citealt{Kobayashi12}). In their hydrostatic burning phase, these stars
lock-up a part of the overall mass and return most of it to the ISM in their AGB phase by stellar winds.
Since the maximum radius of these winds is orders of magnitude smaller than the output range of supernova events  (e.g., radius of Crab remnant: 5.5 Ly (\citealt{Hester08}), while the diameter of the Cat's Eye Nebula is only 0.2 Ly (\citealt{Reed99})), our simulation assumes that stellar winds influence the ISM only in the local calculation cell.
AGB stars provide only a marginal s-process contribution to typical r-process elements like Eu (e.g., \citealt{Travaglio99}). In particular, for this work the s-process contribution to Eu plays a negligible role and we are not considering it here.
\subsubsection{High mass stars (HMS)} \label{HMS}
Massive stars which exceed $8 \text{M}_{\odot}$ are considered to end their life in a core-collapse supernova (CCSN, e.g., \citealt{Thielemann96}, \citealt{Nomoto97}, \citealt{Woosley02}, \citealt{Nomoto13}, \citealt{Jones13}). CCSNe produce most of the O and Mg in the chemical inventory of the galaxy. They provide an important contribution to other $\alpha$-elements (S, Ca, Ti), to all intermediate-mass elements, the iron-group elements and to the s-process species up to the Sr neutron-magic peak (e.g., \citealt{Rauscher02}). Associated to CCSNe, different neutrino-driven nucleosynthesis components might be ejected and contribute to the GCE (e.g., \citealt{Arcones13}, and references therein), possibly including the r-process.
We did not include regular CCSNe as a major source of heavy r-process elements, as recent investigations indicate strongly that the early hopes for a high entropy neutrino wind with the right properties (\citealt{Woosley94}, \citealt{Takahashi94}) did not survive advanced core collapse simulations (e.g., \citealt{Liebendorfer03}) which led to proton-rich environments in the innermost ejecta (see also \citealt{Fischer10}, \citealt{Hudepohl10}), causing rather a so-called $\nu p$-process (\citealt{Frohlich06a}, \citealt{Frohlich06b}, \citealt{Pruet05}, \citealt{Pruet06},  \citealt{Wanajo06}). Further investigations seem to underline this conclusion (recently revisited by \citealt{Wanajo13}), although a more advanced – in medium – treatment of neutrons and protons in high density matter causes possible changes of the electron fraction ($Y_{e}$) of ejecta (\citealt{Martinez-Pinedo12}; \citealt{Roberts12}) and might permit a weak r-process, including small fractions of Eu. Similar effects might be possible via neutrino oscillations (\citealt{Wu14}).
For this reason we did not include regular CCSNe in our GCE simulations, although a weak r-process with small (\citealt{Honda06}-like) Eu contributions could be responsible for a lower bound of [Eu/Fe] observations (see Fig.~\ref{163}), explaining a non-detection of the lowest predicted [Eu/Fe] ratios.
Nucleosynthesis yields for HMS are taken from \cite{Thielemann96} or \cite{Nomoto97}. Assuming a typical explosion energy of $10^{51}$erg, the ejecta are mixed with the surrounding interstellar medium via the expansion of a Sedov-Taylor blast wave, which stops at a radius which contains about $ 5\cdot 10^4 \text{M}_{\odot} $ (see section~\ref{Iteration Procedure} for details on the iteration procedure).
\subsubsection{Supernovae Type Ia (SNIa)} \label{SNIa}
When an IMS is newly born in a binary system, there is a probability that it has a companion in the appropriate mass range leading finally to a SNIa, following a double- or single degenerate scenario. We follow the analytical suggestion of \cite{Greggio05} and reduce the numerous degeneracy parameters to one probability ($P_{SNIa}=9 \cdot 10^{-4}$) for a newly born IMS to actually be born in a system fulfilling the prerequisites for a SNIa. Once the star enters its red giant phase, we let the system perform a SNIa-type explosion and emit the event specific yields (cf. \citealt{Iwamoto99}, model CDD2), which highly enriches the surrounding ISM with iron. For this work we use the same SNIa yields for each metallicity, consistently with the \cite{Argast04} calculations. We are aware that this choice is not optimal, since several SNIa yields including e.g., Mn and Fe depend on the metallicity of the SNIa progenitor (e.g., \citealt{Timmes03}, \citealt{Thielemann04}, \citealt{Travaglio05}, \citealt{Bravo10}, \citealt{Seitenzahl13}). On the other hand, this approximation does not have any impact on our analysis and our conclusions.

\subsubsection{Neutron Star Merger (NSM)} \label{NSM_model}
If two newly born HMS were created in a binary system, they may both undergo a CCSN individually. This could leave two gravitationally bound Neutron Stars (''NS'') behind. Such a system emits gravitational waves and the two NS spiral inwards towards their common center of mass with a coalescence time ($t_{coal}$) until they merge. The actual merging event is accompanied by an ejection of matter and (r-process) nucleosynthesis (\citealt{Rosswog13}, \citealt{Freiburghaus99}, \citealt{Panov08}, \citealt{Korobkin12}, \citealt{Bauswein13}, \citealt{Rosswog13}, \citealt{Rosswog14}, \citealt{Eichler14}, \citealt{Wanajo14}). As all of these publications show the emergence of a strong r-process, in the mass region of Eu they suffer partially from nuclear uncertainties related to fission fragment distributions (see e.g., \citealt{Eichler14}, \citealt{Goriely13}). For our purposes we chose to utilize as total amount of r-process ejecta $ 1.28\cdot 10^{-2} \text{M}_{\odot} $ (consistent with the $1.4 \text{M}_{\odot} + 1.4 \text{M}_{\odot}$ NS collision in \citealt{Korobkin12} and \citealt{Rosswog13}), but distributed in solar r-process proportions, which leads for Eu to a total amount of $10^{-4} \text{M}_{\odot}$ per merger. This value is relatively high in comparison to other investigations in the literature.

Observational constraints for the probability of a newly born star to undergo this procedure ($P_{NSM}$) are provided by e.g., \cite{Kalogera04} who have found a NSM rate of $R_{NSM}=83.0_{-66.1}^{+209.1} \text{Myr}^{-1}$, which corresponds to a $P_{NSM}=0.0180_{-0.0143}^{+0.0453}$.
The coalescence time, $P_{NSM}$ and the event specific yields are important parameters for GCE, and their influence on the GCE are subject of this paper. Concerning the coalescence time scale, it might be more realistic to use a distribution function (e.g., as in \citealt{Ishimaru15}) instead of a fixed value. We utilize this simplified procedure as a  first order approach.

\subsubsection{Magnetorotationally driven supernovae (Jet-SNe)}
A fraction ($P_{Jet-SN}$) of high mass stars end their life as a ''magneto-rotationally driven supernova'' or magnetar, forming in the center a highly magnetized neutron star (with fields of the order $10^{15}$Gauss) and ejecting r-process matter along the poles of the rotation axis (\citealt{Fujimoto06}, \citealt{Fujimoto08}; \citealt{Winteler12}, \citealt{Moesta14}). r-process simulations for such events were first undertaken in 3D by \cite{Winteler12}. For the purpose of this work, we randomly choose newly born high mass stars to later form a Jet-SN. At the end of their life time, they explode similar to a CCSN, however with different ejecta. Based on \cite{Winteler12}, we assume an amount of $14 \cdot 10^{-5} \text{M}_{\odot}$ of europium ejected to the ISM by such an event. In this work, we study the influence of $P_{Jet-SN}$ and the specific Jet-SN yields on the GCE.
\subsection{Observed stellar abundances}
Data for the observed stars to compare our simulation results with are taken from the SAGA (Stellar Abundances for Galactic Archaeology) database (e.g., \citealt{Suda08}, \citealt{Suda11}, \citealt{Yamada13}; in particular [Eu/Fe] abundance observations are mainly from e.g., \citealt{Francois07}, \citealt{Simmerer04}, \citealt{Barklem05}, \citealt{Ren12}, \citealt{Roederer10}, \citealt{Roederer14a}, \citealt{Roederer14b}, \citealt{Roederer14c}, \citealt{Shetrone01}, \citealt{Shetrone03}, \citealt{Geisler05}, \citealt{Cohen09}, \citealt{Letarte10}, \citealt{Starkenburg13}, \citealt{McWilliam13}). From the raw data, we excluded carbon enriched metal poor stars (''CEMPs'') and stars with binary nature, since the surface abundances of such objects are expected to be affected by internal pollution from deeper layers or pollution from the binary companion.
\section{RESULTS}\label{Results}
For a general understanding of the effects of Jet-SNe and NSM on GCE, namely the parameters $P_{NSM}$, $t_{coal}$ and $P_{Jet-SN}$, we performed a number of simulations described in detail below.
\subsection{Coalescence time scale and NSM probability} \label{NSM}
\begin{figure}
\includegraphics[width=84mm]{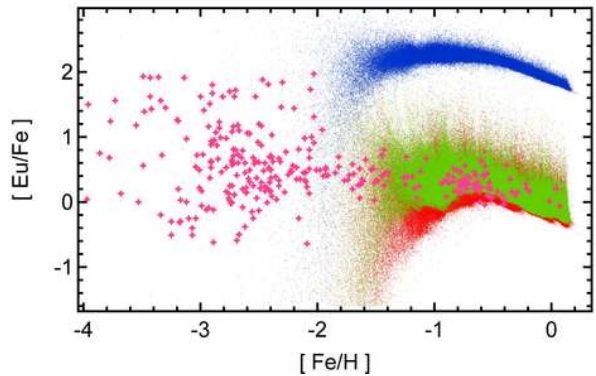}
\caption{Influence of coalescence time scale and NSM-probability on Eu-Abundances in GCE. Magenta stars represent observations. Red dots correspond to model star abundances as in Argast et al. (2004). The coalescence time scale of this event is $10^8$ years and the probability $P_{NSM}$ is set to $4{\cdot}10^{-4}$. Green dots illustrate the effect on the abundances if the coalescence time scale of NSM is shorter (around $10^6$ years). Blue dots show the abundance change if the probability of HMS binaries to later merge in a NSM is increased to $4{\cdot}10^{-2}$ (cf. subsection~\ref{NSM}).}
\label{nfo}
\end{figure}

As a prerequisite, we studied the influence of both coalescence time and the probability of a binary system to become a NSM. In Figure~\ref{nfo}, we present the evolution of [Eu/Fe] abundances when only NSM contribute to the enrichment. The results can be summarized as follows.
\begin{enumerate}
\item Smaller coalescence time scale leads to an enrichment of europium at lower metallicities. Larger coalescence time scale shifts this to higher metallicities.
\item A higher NSM probability shifts towards a quantitatively higher enrichment combined with an appearence at lower metallicities.
\end{enumerate}
These effects can be explained in the following way.
\begin{enumerate}
\item When binary neutron star systems take longer to coalesce, the time between the CCSN of both stars and the NSM event is longer. The longer this delay time, also further nucleosynthesis events occur in the galaxy during this period, enriching the ISM with metals. Thus, when the NSM event finally takes place, surrounding stars have developed a higher [Fe/H] abundance, shifting the system towards higher [Fe/H] abundances, respectively. This implies an overall europium production shift towards higher metallicities.
\item With more binary systems becoming NSM, the produced europium amount per time step is larger, since every event produces the same amount of r-process elements. This leads to a higher [Eu/Fe] abundance, compared to simulations with lower NSM probability. As the fraction of NSM systems is higher while the CCSN rate is constant, larger amounts of europium are produced, while the surrounding medium evolves regularly. This also leads to a higher abundance of europium at lower [Fe/H]. These effects shift the [Eu/Fe] curve to higher values for the same [Fe/H].
\end{enumerate}
All these results are consistent with the earlier conclusions by \cite{Argast04}, stating that it is extremely difficult to reproduce the observed [Eu/Fe] ratios at metallicities [Fe/H]$<-2.5$ by NSM alone. A potential solution would be that the preceding supernovae which produced the two neutron stars of the merging system mix their ejecta with more extended amounts of the ISM. We utilized the results following a Sedov-Taylor blast wave of $10^{51}$erg, which pollutes of the order $ 5\cdot 10^4 \text{M}_{\odot} $ of ISM until the shock is stopped. \cite{vandeVoort15} assumed (in their standard case) the mixing with more than $10^6 \text{M}_{\odot} $ of ISM (\citealt{Shen15} utilized $ 2\cdot 10^5 \text{M}_{\odot} $ in a similar approach). This produces an environment with a substantially lower [Fe/H] into which the NSM ejecta enter. Thus, it is not surprising that in such a case the Eu enrichment by NSM is setting in at lower metallicities. The higher resolution run shown in Fig.~4 of \cite{vandeVoort15} agrees with our results. Thus, the major question is whether such a very much enlarged mixing with the ISM by almost two orders of magnitude can be substantiated. We will discuss these aspect further in section~\ref{Conlusion and Discussion}

\subsection{Probability of Jet-SNe} \label{Jet-SN}
\begin{figure}
\includegraphics[width=84mm]{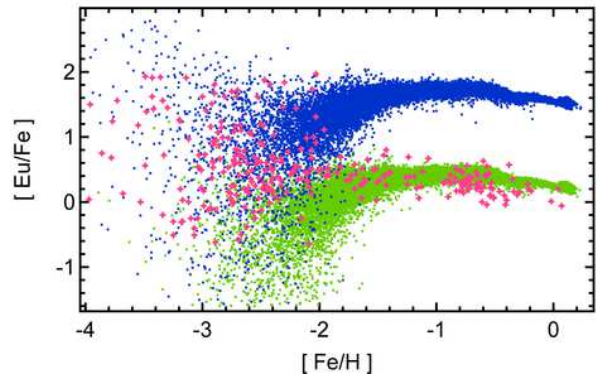}
\caption{Influence of increased Jet-SN probabilities on Eu-Abundances in GCE. Magenta stars represent observations. Green dots represent model star abundances based on Winteler et al. (2012), the Jet-SN probability has been chosen to follow the observations at [Fe/H]$>-1.5$. A good value seems to be $0.1{\%}$~of HMS to end up in a Jet-SN. Note that this model fails to reproduce the observed abundances at lower metallicities. Blue dots illustrate the effect on the abundances if the Jet-SN probability is increased to 1\%. This model better reproduces the observed abundances at lower metallicities, but clearly fails at higher ones.}
\label{Jet-SN1}
\end{figure}

\begin{figure}
\includegraphics[width=84mm]{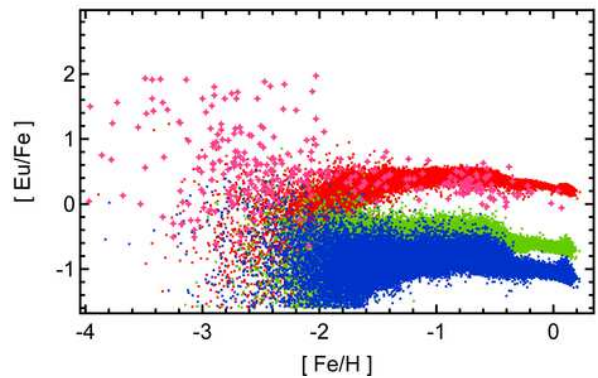}
\caption{Same as Figure~\ref{Jet-SN1}, but with decreased probabilities. Red dots are the same as green dots in Fig.~\ref{Jet-SN1} with Jet-SN probability of $0.1 {\%}$; Green and blue dots represent a Jet-SN probability of $10^{-4}$ and $2\cdot 10^{-5}$, respectively. From the comparison of these models, we can see how decreased event probability shifts the abundance curve down. We also remark an increase of the spread in abundances when the probability is lowered. The rarer a high yield event is, the larger is the spread in abundances.}
\label{Jet-SN2}
\end{figure}

The contribution of Jet-SNe to the galactic Eu abundance differs from that of NSM. Since Jet-SNe explode directly from  a massive star, they contribute much earlier to the chemical evolution than NSM. Since the interstellar matter is distributed more inhomogeneously than in later evolution stages of the galaxy, high [Eu/Fe] abundances are possible in individual stars. This leads to a large spread in the abundances towards lower metallicities. Considering Jet-SNe, the  parameter with the highest impact on GCE for such rare events, similar to NSM events but ''earlier'' in metallicity, is the probability of a massive star to actually become a Jet-SN. A lower probability leads to a smaller overall [Eu/Fe] abundance, while a higher probability leads to larger abundances. However, we also recognize a larger spread in abundances in models with lower probability. This comes from the fact that the high yield of the event only sets an upper limit on the abundances. The rarer an event is, the more and the longer stars remain unpolluted. This results in a larger spectrum of abundances in stars and therefore in a larger spread in [Eu/Fe] ratios.
Note from Fig.~\ref{Jet-SN1} and Fig.~\ref{Jet-SN2} that Jet-SNe might explain the abundances at low metallicities better than NSM. Thus, while Jet-SNe alone could be an explanation for the lower metallicity observations, there is clear evidence of NSM events and therefore we have to examine the combination of both events. Whether the apparently to high concentration of model stars with low [Eu/Fe] values at metallicities $-3<$[Fe/H]$<2$ in comparison to observations is related to observational bias or whether we require another additional source will be discussed in the following sections.


\subsection{Combination of sites} \label{combination}
\begin{figure}
\includegraphics[width=84mm]{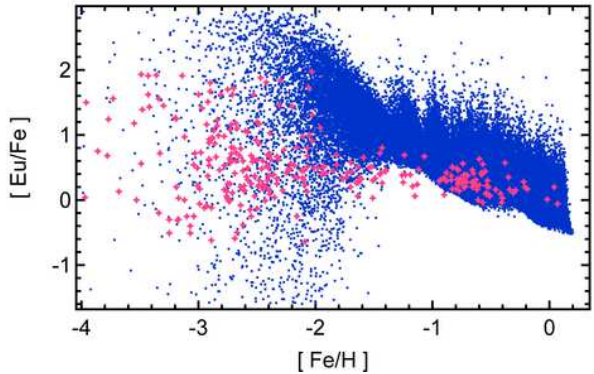}
\caption{Evolution of Eu-abundances in GCE including both Jet-SNe and NSM as r-process sites. Magenta stars represent observations, whereas blue dots represent model stars. Model (Jet+NSM:A) parameters are $P_{NSM}=3.4\cdot 10^{-4}$, $P_{Jet-SN}=0.3\% $, $t_{coal}=1$My.}
\label{23}
\end{figure}

\begin{figure}
\includegraphics[width=84mm]{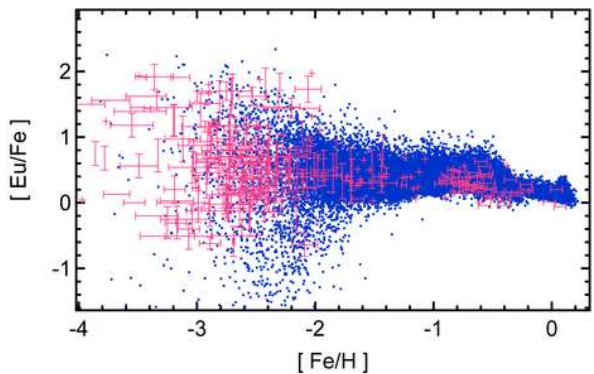}
\caption{Same as Figure~\ref{23}, but with a different parameter set (ModelJet+NSM:B). Magenta stars represent observations (with observational errors; however, magenta stars at low metallicities which carry only horizontal errors represent upper limits). Blue dots represent model stars with $P_{NSM}=3.8\cdot10^{-4}$, $P_{Jet-SN}=0.1\%$, $t_{coal}=10$My.}
\label{163}
\end{figure}

If both sites (Jet-SN and NSM) are considered to contribute to the galactic europium abundances, their contributions overlap. Therefore, parameters which lead to the observed [Eu/Fe] abundances, have to be searched for. As described in section~\ref{NSM}, NSM contribute at a delayed stage to the GCE and in our simulations are unable to reproduce europium abundances at metallicities [Fe/H]$<-2.5$, Jet-SNe, however, contribute europium early, but only in those regions where they occured, and cause a larger spread in the [Eu/Fe] values (cf. section~\ref{Jet-SN}). We have to test whether it is possible to use the same parameters as in sections~\ref{NSM} and \ref{Jet-SN}, since the full combination of both events could lead to an overproduction of elements. We can use the earlier parameter explorations to tune the simulated abundance pattern in order to match the observations.
In the following, we will discuss two possible cases: 
\begin{enumerate}
\item $P_{NSM}=3.4\cdot 10^{-4}$, $P_{Jet-SN}=0.3\% $, $t_{coal}=1$My (hereafter model Jet+NSM:A). The results for the model Jet+NSM:A in comparison with observations are shown in Figure~\ref{23}. This model provide a reasonable explanation of the observations at lower and higher metallicities, but there is an overproduction of europium between $-2<$ [Fe/H] $<-1$. We conclude that larger coalescence time scales and larger probabilities are necessary regarding NSM, and lower probability of Jet-SNe is necessary to flatten and lower the modelled abundance curve.
\item $P_{NSM}=3.8\cdot10^{-4}$, $P_{Jet-SN}=0.1\%$, $t_{coal}=10$My (Model Jet+NSM:B). The results for the model Jet+NSM:B in comparison with observations are shown in Figure~\ref{163}. This model explains the main features of the abundance curve quite well: The spread at low metallicities, the first confinement of the spread at [Fe/H]${\approx}-2$, the plateau between [Fe/H]${\approx}-2$ and [Fe/H]${\approx}-0.6$, and the second confinement of the spread at [Fe/H]$\approx -0.2$. However, there still seem to be difficulties at [Fe/H]${\approx}-2$: the scatter in abundances towards low [Fe/H] ratios seems to be a bit too broad. This spread might be slightly reduced by additional mixing terms (e.g., spiral arms mixing) or an additional source providing ratios of [Eu/Fe]$=-1$, which we did not consider in this work.

\end{enumerate}


Considering Fig.~\ref{23} and Fig.~\ref{163}: While the results from both models Jet+NSM:A and Jet+NSM:B can reproduce the observed spread of [Eu/Fe] in the early galaxy, model Jet+NSB:B seems to better fit the overall [Eu/Fe] vs. [Fe/H] distribution. On the other hand, the evolution of the [Eu/Fe] ratio at low metallicity depends on the r-process production and on the Fe production in CCSNe (see Section~\ref{inhom} and discussion), In Fig.~\ref{23,163euh}, we compare the results for the \emph{enrichment history} of europium in the galaxy according to Jet+NSM:A and Jet+NSM:B models with observations. While the [Eu/H] vs. [Fe/H] ratios predicted by model Jet+NSM:B are in agreement with the observations, model Jet+NSM:A seems to be ruled out.
\begin{figure}
\includegraphics[width=84mm]{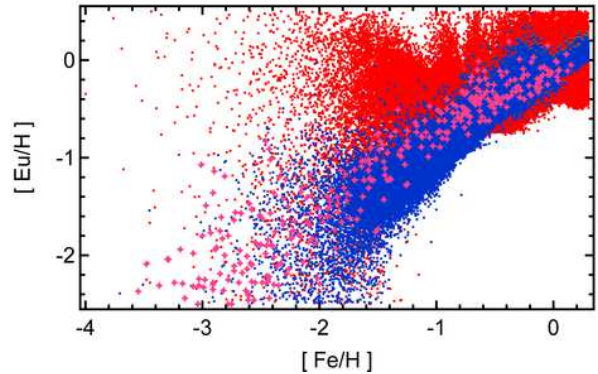}
\caption{Enrichment history for models Jet+NSM:A and Jet+NSM:B (cf. Figure~\ref{23} and \ref{163} for evolution plots). Magenta stars represent observations, whereas blue dots represent model stars as per Figure~\ref{163} (Model Jet+NSM:B). Red dots representing the enrichment history of the simulation as per Figure~\ref{23} (Model Jet+NSM:A) do not suit the observational data.}
\label{23,163euh}
\end{figure}

\section{The importance of inhomogeneities} \label{inhom}
\subsection{Inhomogeneities in GCE}
\begin{figure}
\includegraphics[width=84mm]{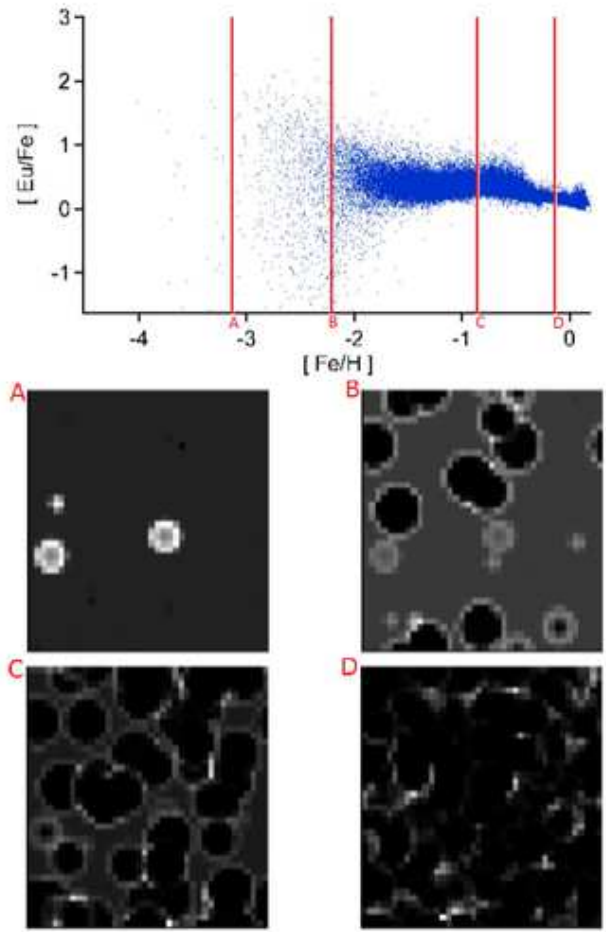}
\caption{The top panel shows the same GCE-model as in Fig.~\ref{163} (Model Jet+NSM:B), but without observations; The red markers refer to the position where a density determination cut through the the xy-plane of the simulation volume is performed. The middle and lower two panels show the density distribution through these planes. The middle left panel corresponds to the very left marker (''A'') position's density profile (approximately 180 million years (My) have passed in the simulation), the middle right panel to the second marker ''B'' ($\approx 290$ My), the lower left panel to the third marker ''C'' ($\approx$ 2 Gy) and the lower right panel to the very right marker ''D'' ($\approx$ 12 Gy).}
\label{ih}
\end{figure}

From observations of [Eu/Fe] in the early galaxy, one of the main features is a spread in the abundance ratios. Our model is able to reproduce these spreads, mainly because of the inhomogeneous pollution of matter. In Fig.~\ref{ih}, we try to illustrate the effect of applying such an inhomogeneous model. For this purpose, we perform a cut through the xy-plane of the simulation volume for specific time steps. These time steps are marked in the top panel of Fig.~\ref{ih}, in order to provide the reader with a quick glance of the extent of the inhomogeneous element distribution at the correspondent metallicities. For each marker, we provide the complete density field at this specific time step in the middle and lower panels (cf. figure caption for details). We show the extent of inhomogeneities in the middle left panel, for the first marker in the upper panel of the Figure. At this time step, we can see - by counting the ''bubble''-style patterns - that at least three supernovae must have taken place before the snapshot. Since such events give rise to nucleosynthesis, the abundances of metals in such a supernova remnant bubble are higher than outside such a remnant. A star being born \emph{inside} such a remnant will inherit more metals than a star born \emph{outside}. Therefore, in the early stages of galactic evolution the stellar abundances are strongly affected by the location \emph{where} a star is born.
Considering much later stages of the evolution, (e.g., the lower right panel of Fig.~\ref{ih}, corresponding to the fourth marker of the upper panel) the supernova remnants have a large overlap. Numerous supernova explosions, have contributed lots of nucleosynthesis all over the galaxy. This leads to an averaged distribution of abundances, including different events and an integral over the initial mass function of stars. Therefore, it resembles a ''mixed'' phase of galactic evolution, where the elements have been homogenised over the whole volume. At this stage of the evolution, it seems not to be so relevant \emph{where} a star was born. As a consequence, there are smaller differences in the abundance of metals in stars. Therefore, a \emph{confinement} in the spread of abundances of chemical elements at later stages of the chemical evolution is obtained. Becoming more and more homogeneous, the [Eu/Fe] value converges to a value that can be obtained by integrating the event yields over the whole IMF.

\subsection{Instantaneous Mixing Approximation}
\begin{figure}
\includegraphics[width=84mm]{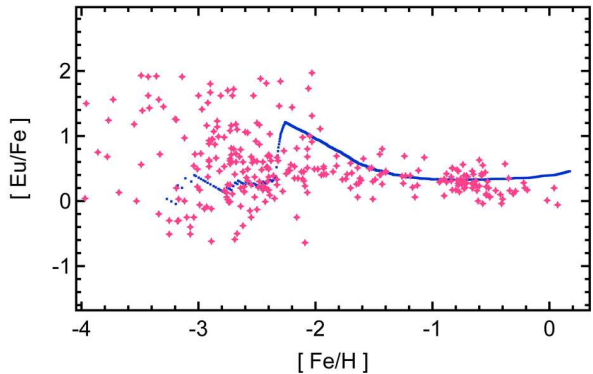}
\caption{Same GCE-model as in Fig.~\ref{163} (Model Jet+NSM:B); however I.M.A. is applied instead of inhomogeneous evolution. One is able to observe a trend in the abundance evolution, however the scatter in the abundance pattern is not present anymore (cf. Section~\ref{inhom} for further discussion). The kink at about [Fe/H]$=-2.5$ is related to the delayed time when NSMs set in and contribute to Eu as well. This Figure can also be compared to Fig.~2 in Matteucci et al. (2014) which shows the contribution of NSMs alone for various merger delay times and Eu production yields and Fig.~5 in Vangioni et al. (2015) [mergers alone being indicated by black lines]. Thus, also in this approach it is evident that the explanation of [Eu/Fe] at low(est) metallicities by NSM alone is not possible.}
\label{163hom}
\end{figure}

A number of recent chemical evolution studies revoked the ''instantaneous mixing approximation'' (I.M.A., e.g., \citealt{Chiappini01}, \citealt{Recchi01}, \citealt{Spitoni09}). The I.M.A. simplifies a chemical evolution model in terms of mass movement. In detail, all event outputs are expected to mix with the surrounding ISM instantaneously. Such approaches always result in an average value of element ratios for each [Fe/H].
Therefore, the I.M.A. scheme all stars at a given time inherit the same abundance patterns of elements and it is impossible to reproduce a scatter in the galactic abundances, which seems to be a crucial ingredient at low metallicities. Indeed, instead of a spread of distributions only one value is obtained for each metallicity. We calculate the best fit model (Jet+NSM:B, cf. Fig.~\ref{163}) with I.M.A. The result can be found in Fig.~\ref{163hom}. The I.M.A. approach may be used to study the chemical evolution trends with a lower computational effort, but Figure~\ref{ih} shows that the reproduction of spreads in abundance ratios due to local inhomogeneities requires to use more complex codes as e.g., the ICE code adopted for this work.

\begin{figure}
\includegraphics[width=84mm]{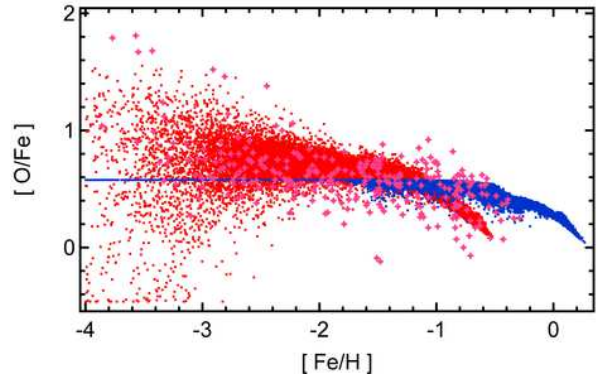}
\caption{Same GCE-model as in Fig.~\ref{163} (Model Jet+NSM:B); Red dots show the abundance evolution of Oxygen when Nomoto et al. (1997) yields are employed, while blue dots represent a far narrower spread at low metallicities if \emph{ad hoc} yields are applied (which still would need to  be optimized to obtain a better agreement with the metallicity evolution between $-1< [\text{Fe}/\text{H}]<0$). Note that the downturn at high metallicities is shifted to higher [Fe/H] values. This is probably due to an overestimate of the total IMF-integrated Fe-production, which should be improved with realistic self-consistent explosion models and their iron yields. While the delay time scale for SNIa is unchanged, earlier CCSN produce more iron, thus dispersing the whole abundance curve. Here we only want to show how changes to possibly more realistic, progenitor-mass dependent, explosion energies can improve the [$\alpha$/Fe] spread, while the [r/Fe] spread is conserved. (Cf. Section~\ref{inhom} for further discussion.)}
\label{527.511_O}
\end{figure}

\begin{figure}
\includegraphics[width=84mm]{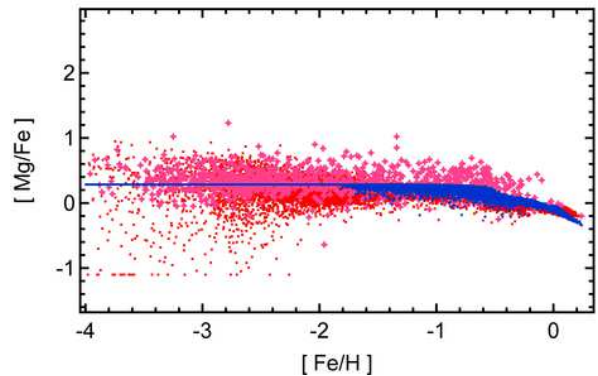}
\caption{Same consideration as in Fig.~\ref{527.511_O}, however with magnesium instead of oxygen. GCE-model as in Fig.~\ref{163} (Model Jet+NSM:B); Red dots show the abundance evolution of magnesium when Thielemann et al. (1996) / Nomoto et al. (1997) yields are employed, while blue dots represent a far narrower spread at low metallicities if \emph{ad hoc} yields are applied. Cf. Section~\ref{inhom} for further discussion.)}
\label{527.512_mg}
\end{figure}

While inhomogeneus GCE codes can explain the spread in r-process elements, there is the question whether they might predict a far too large spread for other elements (e.g., $\alpha$ elements) at low metallicities (with present stellar yields from artificially induced CCSN explosion models). Such effects can also be seen in Fig.~1 in \cite{vandeVoort15} for [Mg/Fe], spreading by more than 1 dex, while observations seem to show a smaller spread up to $0.5$ dex. This can be related to the amount of supernova ejecta being mixed with the ISM (see discussion above and in section~\ref{Conlusion and Discussion}: a more extended mixing reduces this spread), but it can also be related to the supernova nucleosynthesis yields which were never tested before in such inhomogeneous GCE studies.
From general considerations of chemical evolution studies, it is found that there are large uncertainties for GCE studies, particularly the influence of stellar yields (e.g., \cite{Romano10}). In Fig.~\ref{527.511_O}, we show the results of model Jet+NSM:B, using the CCSN yields from from \cite{Nomoto97} and \cite{Nomoto06}, which confirms a large spread in [O/Fe], similar to \cite{vandeVoort15} for [Mg/Fe].  However, present supernova yields are the result of artificially induced explosions with constant explosion energies of the order of $10^{51}$ erg. If we consider that explosion energies might increase with the compactness of the stellar core (i.e., progenitor mass, e.g., \citealt{Perego15}), the heavier $\alpha$-elements and Fe might be enhanced as a function of progenitor mass. On the other hand O, Ne, and Mg yields are dominated by hydrostatic burning and also increase with progenitor mass (e.g., \citealt{Thielemann96}). This could permit to obtain more constant $\alpha$/Fe ratios over a wide mass range, although the total amount of ejecta differs (increases) as a function of progenitor mass. This scenario does not take into account all the complexity and the multi-dimensional nature of the CCSN event (e.g., \citealt{Hix14}, and references therein) that should be considered, but it may be interesting to test its impact in our GCE simulations. In Figs.~\ref{527.511_O}~and~\ref{527.512_mg}, we show the results for tests where we:
\begin{enumerate}
\item replace the \cite{Nomoto97} iron yields by \emph{ad-hoc} yields, fitting, however, the observed SN1987A iron production;
\item keep the same CCSN rate as in the previous models;
\item adopt the parameters to study the r-process nucleosynthesis of Model Jet+NSM:B, obtaining the same [Eu/H] ratio.
\end{enumerate}
This leads, based on the adopted CCSN yields, to a possibility to minimize the spread in $\alpha$-elements at low metallicities, while keeping the spread in the r-process element evolution. Therefore, the spread of [O/Fe] obtained from GCE simulations for the early galaxy is strongly affected by the uncertainties in the stellar yields, and it is difficult to disentangle them from more intrinsic GCE uncertainties. This means that at this stage it is not obvious whether an overestimation of the observed [O/Fe] spread is a problem of the ICE code, the observations could rather provide a constraint on stellar yields. In particular, the use of realistic, self-consistent, explosion energies, might reduce the spread at low metallicities to a large extent.
Another fundamental point is related to the discussion in Section~\ref{Results} concerning [Eu/Fe]. At this stage, we consider [Eu/H] as more constraining to study the r-process nucleosynthesis compared to [Eu/Fe], since Fe yields from CCSNe are affected by large uncertainties. Therefore, the model Jet+NSM:B is recommended compared to Jet+NSM:A (see also Fig.~\ref{23,163euh}).


\section{Conclusion and discussion}\label{Conlusion and Discussion}
The main goal of this paper was to reproduce the solar europium abundance as well as the evolution of  [Eu/Fe] vs. [Fe/H] throughout the evolution of the galaxy.
For this reason we have studied the influence of two main r-process sites (NSM and Jet-SNe) on the GCE. 

Our simulations were based on the inhomogeneous chemical evolution (ICE) model of \cite{Argast04}, with updated nucleosynthesis input for the two sites considered, their respective occurrence frequencies / time delays, and a model resolution of $(50\textit{pc})^3$. The main conclusions are that:

\begin{enumerate}

\item The production of heavy r-process matter in NSM is evident since many years (see \citealt{Freiburghaus99} and many later investigations up to \citealt{Korobkin12}, \citealt{Rosswog13}, \citealt{Bauswein13}, \citealt{Rosswog14}, \citealt{Just14}, \citealt{Wanajo14}, \citealt{Eichler14}, \citealt{Mendoza15}). Our implementation of NSM in the inhomogeneous chemical evolution model ''ICE'' can explain the bulk of Eu (r-process) contributions in the galaxy for [Fe/H]$>-2.5$, but have problems to explain the amount and the spread of [Eu/Fe] at lower metallicities. This is in agreement with the initial findings of \cite{Argast04}. Recent SPH-based studies by \cite{vandeVoort15} make use of a mixing of the ejecta with $3 \cdot 10^6 M_\odot$, a further study by \cite{Shen15} utilizes a mixing with $2 \cdot 10^5M_\odot$ up to $8 \cdot 10^5M_\odot$.
The mixing volume we utilize, based on the Sedov-Taylor blast wave approach, would be related to a subgrid-resolution in these studies, but this treatment is essential for the outcome. Mixing initially with a larger amount of matter causes smaller [Fe/H] ratios into which the r-process material is injected.

We have tested such differences in mixing volumes/masses also within our ICE approach. Fig.~\ref{Sweepup} shows the results we obtain when changing from the Sedov-Taylor blast wave approach to a mixing mass of $2 \cdot 10^5 M_\odot$ (like in \citealt{Shen15}), and we can see that we essentially reproduce their results. On the other hand, a higher resolution test in section 3.1 of \cite{vandeVoort15} is essentially in agreement with our results presented above. Thus, these differences are not based on the differences in sophistication of the multi-D hydrodynamics approach, permitting to model energy feedback from supernovae, outflows and infall, they can rather be linked directly to the mixing volumes of supernova ejecta. This requires further studies in order to understand whether there exist physical processes (on the timescale of the delay between the supernova explosions and the merger event) which permit a mixing beyond the Sedov-Taylor blast wave approach.

\begin{figure}
\includegraphics[width=84mm]{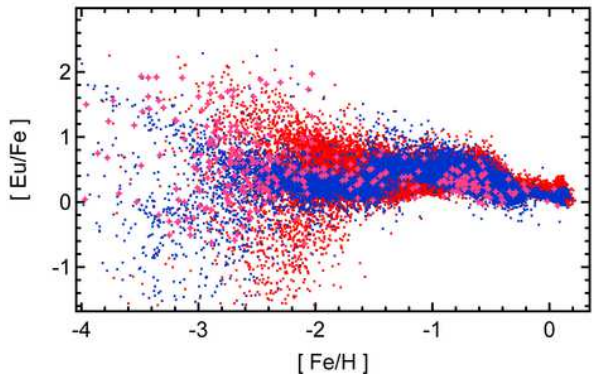}
\caption{Effect of slightly increased sweep-up mass on GCE. Magenta stars represent observations. Red dots show model stars as per our reference model JET+NSM:B. Blue dots represent a model where every CCSN pollutes $2\cdot 10^5 M_{\odot}$ of ISM. The dominant effect of this increased sweep up mass is to decrease the scatter of abundances and to shift the abundance curve towards lower metallicities.)}
\label{Sweepup}
\end{figure}

\item The production of heavy r-process elements in a rare species of CCSNe with fast rotation rates and high magnetic fields, causing (fast) jet ejection of neutron-rich matter along the poles has first been postulated by \cite{Cameron01}. This was followed up in rotationally symmetric 2D calculations \cite{Fujimoto06}, \cite{Fujimoto08} and the first 3D calculations by \cite{Winteler12}. These calculations still depend on unknown rotation velocities, and magnetic field configurations before collapse, however, they agree with the observations of magnetars and neutron stars with magnetic fields of the order $10^{15}$Gauss which make up about $1\%$ of all observed neutron stars. Further 3D calculations by \cite{Moesta14} and recent 2D calculations by \cite{Nishimura15}, might indicate that not all events leading to such highly magnetized neutron stars are able to eject the heaviest r-process elements in solar proportions. Thus, probably less than 1\% of all CCSNe end as Jet-SNe with a full r-process.

When introducing Jet-SNe with ejecta as predicted by \cite{Winteler12}, they can fill in the missing Eu at lower metallicities and reproduce the spread in [Eu/Fe], in agreement with the recent findings of \cite{Cescutti14}. We find that a fraction of $0.1\%$ of all CCSNe which end up in this explosion channel provides the best fit. This would mean that not all but only a fraction of magnetar events which produce the highest magnetic field neutron stars are able to eject a main r-process composition of the heaviest elements, as discussed above.

Our conclusion is that both sites acting in combination provide the best scenario for understanding [Eu/Fe] observations throughout galactic history, with typical probabilities for NSM formations and (merging) delay times as well as probabilities for Jet-SNe.
As a side effect we realized that present supernova nucleosynthesis yield predictions, based on induced explosions with a single explosion energy throughout the whole mass range of progenitor stars, bear a number of uncertainties. While apparently too large scatters of alpha/Fe ratios can be obtained in inhomogeneous chemical evolution models, when utilizing existing nucleosynthesis predictions from artificial explosions with energies of 1 Bethe, this might not be due to the chemical evolution model. Such deficiencies can be cured by assuming  larger mixing masses with the ISM for supernovae explosions (\citealt{vandeVoort15}, \citealt{Shen15}), or the introduction of an artificial floor of abundances based on IMF-Integrated yields of CCSNe for metallicities at [Fe/H]$=-4$, but it could in fact just be due the non-existence of self-consistent CCSN explosion models. We have shown that an explosion energy dependence on the compactness of the Fe-core, related to the main sequence mass, could solve this problem as well by modifying the nucleosynthesis results. Therefore, self-consistent core collapse calculations with explosion energies varying with progenitor mass and possibly other properties like rotation are highly needed. Although we have obtained a good accordance with the observed Eu abundances, the true origin of r-process elements might still require additional insights. The present investigation may be used to put constraints on the yields, as well as essential properties and occurrence frequencies of sites.

There exist a number of open questions not addressed in the present paper, related (a/b) to production sites and (c) to the true chemical evolution of the galaxy.
\begin{enumerate}

\item As discussed in detail in section~\ref{HMS}, we did not include ''regular'' CCSNe from massive stars as contributors to the main or strong r-process, producing the heaviest elements in the Universe. However, as already mentioned in section~\ref{HMS}, there exists the chance for a weak r-process, producing even Eu in a Honda-style pattern in such events. This could provide the correct lower bound of [Eu/Fe] in Fig.~\ref{163} and would be consistent with the recent findings of \cite{Tsujimoto14}.

\item We did not include the effect of NS-BH mergers in the present paper. They would result in similar ejecta as NS-NS mergers per event (\citealt{Korobkin12}), but their occurrance frequencies bear high uncertainties (\citealt{Postnov14}). \cite{Mennekens14} provide a detailed account of their possible contribution and also discuss their contribution to global r-process nucleosynthesis. One major difference with respect to our treatment of NSM in chemical evolution relates to the fact, that (if the black hole formation is not causing a hypernova event but rather occurring without nucleosynthesis ejecta) only one CCSN is polluting the ISM with Fe before the merger event, in comparison to two CCSNe. This would lead to a smaller [Fe/H] ratio in the ISM which experiences the r-process injection, and just to an ''earlier'' appearance of high [Eu/Fe] ratios in galactic evolution. If we assume that BH formations are as frequent as supernova explosions, an upper limit of the effect would be that all NS-NS mergers are replaced by BH-NS mergers, moving the [Eu/Fe] features to lower metallicties by a factor of $2$. However, the lower main sequence mass limit for BH formation is probably of the order $20M_\odot$, and only a small fraction of core collapses end in black holes. Therefore, we do not expect that the inclusion of NS-BH mergers shifts the entries by more than $0.15$ in Fig.~\ref{nfo}. This by itself would not be a solution in terms of making only compact (i.e. NS-NS and NS-BH) mergers responsible for the r-process at very low metallicities.

\item There have been suggestions that the Milky Way in its present form resulted from merging subsystems with a different distribution of masses. Such “dwarf galaxies” will experience different star formation rates. It is known that different star formation rates can shift the relation [X/Fe] as a function of [Fe/H]. If the merging of such subsystems will be completed at the time when type Ia supernovae start to be important, the relation  [X/Fe]=f([Fe/H]) will be uniform at and beyond [Fe/H]$>-1$, but it can be blurred for low metallicities between the different systems, possibly leading also to a spread of the onset of high [Eu/Fe] ratios at low metallicities (\citealt{Ishimaru15}). The result depends on the treatment of outflows, should in principle be tested in inhomogeneous models, and also already be present in the simulations of \cite{vandeVoort15} and \cite{Shen15}. But it surely requires further investigations to test fully the impact of NSM on the r-process production in the early galaxy.
\end{enumerate}
\end{enumerate}

Future studies will probably require a distribution of delay times for NSM events, a test of the possible contributions by BH-NS mergers, a better understanding of yields, and improvements in understanding mixing processes after supernovae explosions and during galactic evolution. Testing the full set of element abundances from SNe Ia and CCSNe as well as the two sources discussed above, in combination with extended observational data, will provide further clues to understanding the evolution of galaxies.

\section{Acknowledgements}
The authors thank Dominik Argast for providing his GCE code ''ICE'' for our investigations.
MP thanks the support from the Swiss National Science Foundation (SNF) and the "Lend\"ulet-2014" Programme of the Hungarian Academy of Sciences (Hungary).
BW and FKT are supported by the European Research Council (FP7) under ERC Advanced Grant Agreement No. 321263 - FISH, and the Swiss National Science Foundation (SNF). The Basel group is a member in the COST Action New Compstar.
We also thank Almudena Arcones, John Cowan, Gabriel Martínez-Pinedo, Lyudmila Mashonkina, Francesca Matteucci, Tamara Mishenina, Nobuya Nishimura, Igor Panov, Albino Perego, Tsvi Piran, Nikos Prantzos, Stephan Rosswog, and Tomoya Takiwaki for helpful discussions during the Basel Brainstorming workshop; Camilla J. Hansen, Oleg Korobkin, Yuhri Ishimaru, and Shinya Wanajo for discussions at ECT* in Trento, and Freeke van de Voort for providing details about their modelling.
We would also like to acknowledge productive cooperation with our Basel colleagues Kevin Ebinger, Marius Eichler, Roger K\"appeli, Matthias Liebend\"orfer, Thomas Rauscher, and Christian Winteler. Finally, we would also like to thank an anonymous referee who provided useful and constructive suggestions and helped improving the paper.

\end{document}